\documentclass[authoryear,12pt,nopreprintline]{elsarticle}

\usepackage{amsmath}
\usepackage{graphicx}   
\usepackage{float}
\usepackage{siunitx}
\usepackage{miller}     
\usepackage{soul}       
\usepackage{makecell}   
\usepackage{booktabs}   
\usepackage{lineno}

\begin{document}

\begin{frontmatter}

\title{Microstructure sensitive recurrent neural network surrogate model of crystal plasticity}
\date{\today}

\author[ukaea]{Michael D. Atkinson\corref{cor1}}
\ead{michael.atkinson@ukaea.uk}
\cortext[cor1]{Corresponding author}
\author[ukaea]{Michael D. White}
\author[ukaea]{Adam J. Plowman}
\author[ukaea]{Pratheek Shanthraj}
\affiliation[ukaea]{organization={UK Atomic Energy Authority},
            addressline={Culham Campus},
            city={Abingdon},
            postcode={OX14 3DB},
            country={UK}}



\begin{abstract}
The development of next-generation structural materials for harsh environments requires rapid assessment of mechanical performance and its dependence on microstructure. While full-field crystal plasticity (CP) models provide detailed insights, the high computational cost limits their use with uncertainty quantification workflows and in component-scale simulation. Surrogate models based on recurrent neural networks (RNNs) have shown promise in reproducing history-dependent mechanical behaviour but are applied to models with either fixed microstructure or representative volume elements. Here, we develop a microstructure sensitive RNN surrogate that predicts homogenised stress responses directly from three-dimensional grain structures and arbitrary deformation histories. The architecture employs a gated recurrent unit (GRU) with mappings from microstructure to both the initial hidden state and sequence inputs, allowing the model to capture path dependence and microstructure variability. Training data comprised of over \num{300000} CP simulations generated using combinations of randomly generated microstructures and loading paths. The model was found to reproduce CP predictions for both in-distribution validation data and unseen deformation modes, with errors of \SIrange{2}{3}{\mega\pascal}. Out-of-distribution microstructures were more difficult to predict, emphasising the need for representative training data with, for example, heavily textured microstructures. Embedding the model into a multiscale framework demonstrates its ability to replace conventional constitutive updates, reducing computational cost while preserving key features of the stress distribution. These results establish microstructure-informed RNN surrogates as a promising alternative to direct CP simulations, offering a pathway toward rapid multiscale modelling and uncertainty quantification.
\end{abstract}

\end{frontmatter}

\section{Introduction}
The mechanical performance of polycrystalline metals is strongly dependent on their microstructure, which influences both the local behaviour by the heterogeneous distribution of strain and the average behaviour at a larger scale. Understanding structure-property relationships is a central goal of materials science to enable the development of materials for use in new environments. This knowledge can be gained experimentally through large testing campaigns, but the cost and time requirements mean that simulation using materials models at various length scales is used to supplement experimental data. At the microstructure scale, full-field crystal plasticity (CP) models developed over the last few decades are now routinely used to understand deformation behaviour in specific cases \citep{Dunne2007,Lim2011,Guan2016,Yuan2019}. These continuum models explicitly include microstructure by discretising in the form of a finite element mesh (CPFEM \citep{Kalidindi1992,Marin1998a,Roters2010a}) or 3D images linked to efficient fast Fourier transform based solution schemes (CPFFT \citep{Lebensohn2012,Roters2018}). Despite the increase in compute power since their early development, these models are still prohibitively expensive to be directly included in simulations at the scale of components or uncertainty quantification workflows.

Surrogate modelling or model order reduction is a powerful technique used to reduce the computation demand of complex models and allow them to be applied in scenarios that would not otherwise be possible. Surrogate models act as fast-executing approximations of the underlying physical model, typically using data from a set of simulations to produce a response surface parametrised over the inputs of interest. Gaussian process regression and neural networks are popular techniques used for application across materials science and engineering \citep{bishara_state---art_2023, Tran2021, kusampudi_inverse_2023, dorward_calibration_2024}. CP models use internal state variable formations to capture the evolving mechanical properties of the material as a function of the microstructure features, leading to a history dependent model. This history must be considered for general loading conditions to be modelled, and recurrent neural networks (RNNs) have been shown to perform well in reproducing the response of CP models with fixed microstructure. 

RNNs are used to model time series data by sequentially updating an internal state with input data from each point in the series and projecting the state to an output vector \citep{hochreiter_long_1997, cho_properties_2014}. An RNN surrogate can be used to solve a homogenisation problem where a variable loading path is applied to a fixed microstructure and the stress response of the ensemble is output. Gated RNN architectures have been demonstrated for this problem, including long short-term memory (LSTM) networks applied to plasticity in heterogenous composites \citep{ghavamian_accelerating_2019} and viscoelasticity \citep{chen_recurrent_2021}, and gated recurrent unit (GRU) networks applied to various metal plasticity models \citep{abueidda_deep_2021,tancogne-dejean_recurrent_2021,gorji_potential_2020}. These ``off-the-shelf'' networks have the advantage of being included in machine learning frameworks (e.g. PyTorch \citep{paszke_pytorch_2019}), so are easy to implement with efficient computation. \citet{bonatti_importance_2022} proposed an RNN architecture tailored to micromechanics applications, with a minimal representation and ensuring that the loading path is not dependent on the discretisation to a time series. They demonstrated this architecture to produce a surrogate of a CP model with fixed microstructure under monotonic loading conditions \citep{bonatti_cp-fft_2022}. \citet{liu_learning_2023} go on to formulate the recurrent neural operator, a time continuous generalisation of an RNN. They demonstrate the model with complex load paths including load reversals and abrupt changes of strain rate, applied to both composite materials and 2D CP. Instead of taking a purely data learning approach, constraints motivated by the physical systems being modelled can also be included to improve accuracy, with approaches including adding loss terms to ensure positive dissipation \citep{borkowski_recurrent_2022, rezaei_learning_2024}, mechanical equilibrium \citep{harandi_spifol_2025, keshavarz_advancing_2025} or a consistent simple representation of the hidden state \citep{jones_attention-based_2025}.

In addition to the applied loading, the homogenised response of a material is also influenced by the initial microstructure. Statistical parameters could be used to describe the microstructure with a surrogate defined over a low dimensional parameter space, but the direct mapping between individual microstructures and stress response would be lost. Neural operators have been used as surrogates of physical systems represented by partial differential equations (PDEs) and extensions, such as Fourier neural operators \citep{bhattacharya_learning_2024} and Fourier neural mappings \citep{huang_operator_2025, bhattacharya_learning_2025}, have been used as surrogates with variable starting conditions. The mappings take input image data of the initial state representing, for example, the distribution of phases in a composite and require large training datasets. Convolutional neural networks can be used to compress the image representation, allowing integration into surrogates using feedforward networks with fixed loading \citep{he_material-response-informed_2024, white_3d_2025} or RNNs \citep{frankel_predicting_2019}. Proper orthogonal decomposition (POD) has also been suggested to express model evolutional over a set of highest variability modes in the training set to reduce the number of parameters \citep{vijayaraghavan_data-driven_2023}. In the case of polycrystals, the inherent structure can be included in the microstructure representation using graphs to represent the local dependence between grains that may exist \citep{patel_equivariant_2024, hu_temporal_2024}. \citet{mozaffar_deep_2019} then considers the best approach to integrate microstructure into an RNN, mapping to either the initial hidden state or combined with input data at each step of the sequence.

In this work, we construct an RNN surrogate model for the homogenised response of a 3D full-field CP model. Mappings from a microstructure representation to both the initial hidden state and sequence input are used to create microstructure dependence of small volumes of material. Training data of both the loading paths and initial microstructures are randomly generated, with the aim of creating a surrogate that generalises to any input data. 

\section{Microstructure sensitive model}
The proposed surrogate model aims to produce volume average stress response of a small region of microstructure given an input deformation history and initial microstructure description.
\subsection{Model architecture}
A general recurrent neural network consists of a hidden state vector that is updated at every point of a sequence (or time-step) to learn the underlying process in which the sequence was generated. At each point of the sequence, an input vector ($x_t$) and the previous hidden state ($h_{t-1}$) are used to update the hidden state ($h_t$), and then a transformation is defined to the output of interest ($o_t$). We have used a gated recurrent unit (GRU) to define the RNN \citep{cho_properties_2014}, which consists of two gated inputs of fully connected layers from $x_t$ and $h_{t-1}$ with sigmoid activation (reset gate ($r_t$) and update gate ($z_t$)). The reset gate is then used to define a new hidden state proposal ($\hat{h}_t$), where the gate defines the weight contributed from the previous hidden state. The final updated hidden state is the sum of the proposal and the previous hidden state weighted by the update gate. We use PyTorch \citep{paszke_pytorch_2019} to implement the model, where an efficient GRU network is predefined. The GRU network is defined by,
\begin{equation}
\begin{aligned}
    r_t &= \sigma(W_{ir}x_t + W_{hr}h_{t-1} + b_r) \\
    z_t &= \sigma(W_{iz}x_t + W_{hz}h_{t-1} + b_z) \\
    \hat{h}_t &= tanh(W_{ih}x_t + r_t \odot (W_{hh}h_{t-1}) + b_h) \\
    h_t &= (1-z_t) \odot \hat{h}_t + z_t \odot h_{t-1} ,
\end{aligned}
\end{equation}
where $W_*$ and $b_*$ are the parameters and $\odot$ is the component-wise product.

The hidden state of the RNN represents the internal state of the CP model, which is a function of the fixed initial microstructure and internal variables which vary with the loading applied. Deformation gradient $F$ and the deviatoric part of Cauchy stress $S$ are used to define the time-varying loading applied in the model, which are the sequence inputs and outputs of the GRU block. Initially, the internal state of the CP model is defined by the microstructure and so we take the initial hidden state to be a function of the microstructure ($m$) defined by a neural network mapping ($f$). The memory of the microstructure will decay along the length of the sequence, so a second mapping of the microstructure ($g$) is combined with $F$ to ensure the loading history remains dependent on the microstructure. Both the input ($x_t$) and ($o_t$) are passed through a single fully connected layer. For the input, this combines the contribution of loading and microstructure and for the output this projects the hidden state to the 6 unique components of stress. These projections are defined as,
\begin{equation}
\begin{aligned}
    h_0 &= f(m) \\
    m_0 &= g(m) \\
    x_t &= W_i[F_t, m_0] + b_i \\
    o_t &= \sigma(W_oh_t + b_o).
\end{aligned}
\end{equation}

\begin{figure}
  \centerline{\includegraphics[scale=0.8]{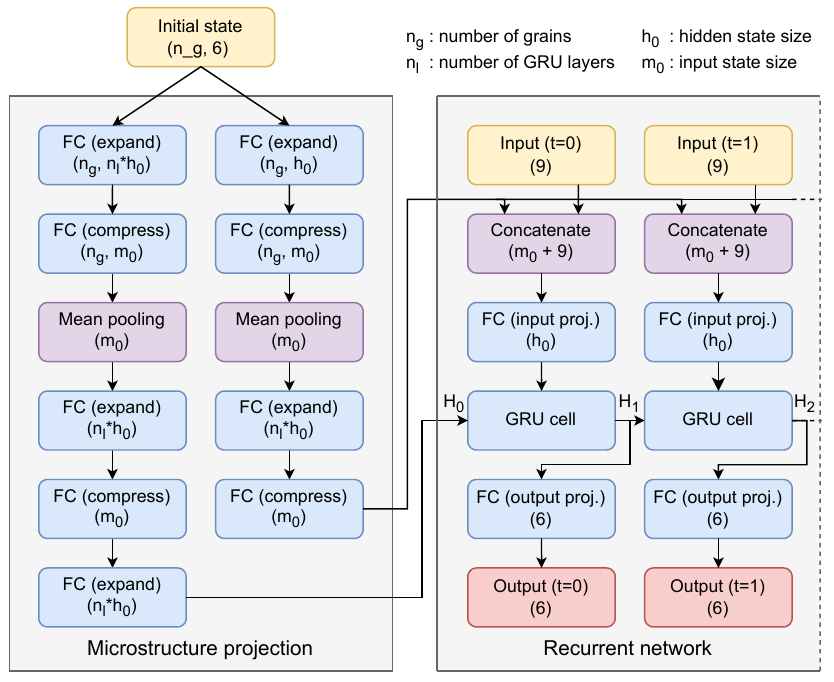}}
  \caption{Network diagram of proposed surrogate model. The left section is a mapping from microstructure parameters (grain locations and orientations) to a state vector and the right the RNN mapping a sequence of average deformations to average stresses.}
  \label{fig:network_diagram}
\end{figure}

Voronoi tessellations are used to represent the microstructure and are defined by a seed position (3 parameters) and crystal orientation (3 parameters) for each grain. There is significant redundancy in this representation as a result of symmetrically equivalent crystal orientations and the swapping of the ordering of the grains leaving the tessellation unchanged. The orientation redundancy is removed by mapping each orientation to the equivalent orientation within the fundamental zone of the crystal structure. A mean pooling operation over the grain level attributes is used in the microstructure mappings ($f$ and $g$) to remove the grain swapping symmetry. A digram of the RNN model with microstructure projections is shown in Figure \ref{fig:network_diagram}.

\subsection{Crystal plasticity model}
The phenomenological CP model within DAMASK \citep{Roters2018} is used to simulate full-field mechanical response of polycrystalline aggregates. DAMASK implements a finite strain continuum model with a multiplicative decomposition of deformation gradient ($\mathbf{F}$) into elastic ($\mathbf{F}^e$) and plastic ($\mathbf{F}^p$) parts. Elastic deformation is governed by Hooke's law and plastic deformation by a viscoplastic flow rule. The flow rule follows the early work of Asaro and Needleman \citep{Asaro1985}, with a power law dependence with a contribution from each of a set of allowed slip systems,
\begin{align}
\label{eqn:flowRule}
\mathbf{L^p} &= \sum_{\alpha}{\dot{\gamma}^\alpha\mathbf{s^\alpha}\otimes\mathbf{n^\alpha}},    &    \dot{\gamma}^\alpha &= \dot{\gamma}_0\text{sgn}\left(\frac{\tau^\alpha}{g^\alpha}\right)\left|\frac{\tau^\alpha}{g^\alpha}\right|^{\frac{1}{m}},
\end{align}
where $\mathbf{L^p}$ is the plastic velocity gradient, $\alpha$ enumerates the possible slip systems, $\mathbf{s^\alpha}$ and $\mathbf{n^\alpha}$ are the slip direction and slip plane normal, $\dot{\gamma}_0$ is a reference shear rate, $m$ is the strain rate sensitivity and $\tau^\alpha$ and $g^\alpha$ are the current and critical resolved shear stress (CRSS) of each slip system. A spectral solver is used to solve for equilibrium at each step of a simulation, allowing efficient computation of the model in a regular cuboid domain with periodic boundary condition. The deformation is imposed by specifying the first component of the Fourier transform (mean) of the deformation gradient field.

The state at each point of a volume element (VE) is described by the crystal orientation and CRSS values ($g^\alpha$), which evolve throughout the simulation leading to a history dependent model. Grain rotation is calculated from a polar decomposition of $\mathbf{F}^e$ which is used to update the crystal orientations at each time step. The slip system hardness is modelled by a Voce type hardening law with latent hardening,
\begin{equation}
\label{eqn:hardening}
\dot{g}^\alpha =  h_0\sum_\beta h^{\alpha\beta}\left|\dot{\gamma}^\beta\right|\left|1 - \frac{g^\beta}{g_s}\right|^a \text{sgn}\left(1 - \frac{g^\beta}{g_s}\right),
\end{equation}
where $h_0$, $g_s$ and $a$ are material parameters and $h^{\alpha\beta}$ the latent hardening matrix. 

Material parameters were taken from the literature and are listed in Table \ref{tab:sim_params_damask}. These represent aluminium with a face-centred cubic crystal structure and the \hkl{111}\hkl<110> slip systems used for plastic deformation. 

\begin{table}
  \caption{Material parameters for the CP simulations.}
  \begin{tabular}{l l l l l l}
    \thead[l]{$m$} & \thead[l]{$\dot{\gamma}_0$ (\si{\per\second})} & \thead[l]{$h_0$ (\si{\mega\pascal})} & \thead[l]{$g_0$ (\si{\mega\pascal})} & \thead[l]{$g_s$ (\si{\mega\pascal})} & \thead[l]{$a$}\\\hline
    \num{0.05} & \num{e-3} & \num{75} & \num{31} & \num{63} & \num{2.25} \\
    \thead[l]{$C_{11}$ (\si{\giga\pascal})} & \thead[l]{$C_{12}$ (\si{\giga\pascal})} & \thead[l]{$C_{44}$ (\si{\giga\pascal})} & \thead[l]{$h^{\alpha\beta}$} & \\\hline
    \num{106.75} & \num{60.41} & \num{28.34} & \multicolumn{3}{l}{\num{1.4} for co-planar, \num{1} otherwise}   
  \end{tabular}
  \label{tab:sim_params_damask}
\end{table}

\subsection{RNN model training}
\begin{figure}
  \centerline{\includegraphics{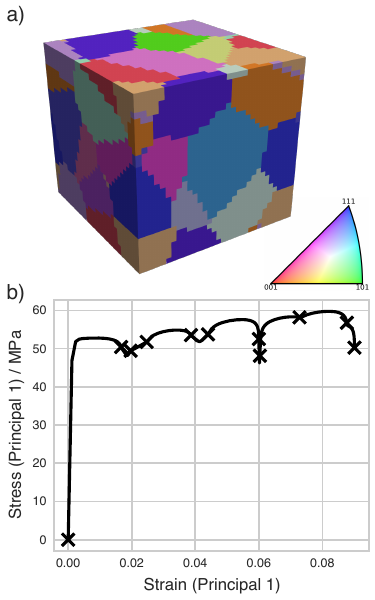}}
  \caption{A typical value from the training set for a) microstructure Voronoi tessellation shown in inverse pole figure (IPF) colouring along a \hkl<100> direction and b) a loading path, where crosses mark the sections of different strain rate.}
  \label{fig:ve_and_loadcase}
\end{figure}

Training data for the RNN model consists of inputs to define a microstructure VE and loading history, which are then used in the CP model described in the previous section to produce the average stress response of a VE. Voronoi tessellation is used to define each VE by a set of seed points and crystallographic orientations. Each seed point is sampled uniformly from the product of three unit intervals $s\sim U(0,1)^3$. Orientations are uniformly sampled from $SO(3)$ and then transformed to the cubic crystallographic fundamental zone using the DAMASK pre-processing library \citep{mentock_python_2025}. This process reduces the input space size for the orientations and ensures only one of the 24 possible symmetrically equivalent orientations is consistently selected. Although the microstructure projection (Figure \ref{fig:network_diagram}) can encode a variable number of grains in a VE, we restrict this work to a fixed \num{20} grains. The Voronoi tessellation is discretised to a regular \numproduct{32x32x32} grid to produce a grain image for input to DAMASK. An example VE is shown in Figure \ref{fig:ve_and_loadcase}a.

Loading histories are generated randomly with the restriction that loading is monotonic and in the same direction throughout but the loading rate can vary. Each load path is defined by a macroscopic deformation gradient $\mathbf{F}$ that varies over time and is applied to the VE in a simulation. The path is first split into \num{10} sections $\mathbf{F}^m$ at the end of each section ($1\leq m\leq10$) is defined by,
$$
\mathbf{F}^m=\mathbf{\tilde{F}}^m(\det\mathbf{\tilde{F}}^m)^{-\frac{1}{3}}, \qquad
\mathbf{\tilde{F}}^m=\mathbf{F}^{m-1}+F^m_{rate}\mathbf{F}_{dir},
\qquad
\mathbf{F}^0=\mathbf{I},
$$
where each component of $\mathbf{F}_{dir}$ is uniformly independently sampled from $U(-1,1)$ and defines the direction of loading, which is fixed throughout the loading path. The loading rate is varied in each section of the load path by $F^m_{rate}$ which is sampled from $U(0,F_{max})$, where $F_{max}$ is the maximum deformation rate of \num{2e-2} here. To prevent the development of large unphysical volumetric stresses due to physical constraint, each target $\mathbf{\tilde{F}}^m$ is scaled to remove the volume change and the surrogate model is trained on only the deviatoric part of stress (referred to as stress throughout for brevity). Piecewise cubic hermite interpolating polynomials (PCHIP) are then used to produce smooth trajectories over \num{200} \SI{1}{\second} time steps. An example load path is shown in Figure \ref{fig:ve_and_loadcase}b. 

A dataset is then produced by running CP simulations for a set of generated VEs and load paths. Although these two parameters could be sampled individually for each simulation, here we use a structured approach so that more than one load path is simulated for each VE, and vice versa. First, a matrix of \num{2000} VEs and \num{5000} load paths is generated and then simulations are defined by randomly selecting points from this matrix. A high performance compute (HPC) cluster is used to run the CP simulations in large batches, allocating 8 CPU cores per simulation. This produced a dataset of \num{328386} simulations, which is divided into training and validation sets in the ratio 0.94:0.06. The data is normalised by individually scaling each component of the tensor parameters to the range \numrange{0}{1} by the minimum and maximum value of the component in the dataset. Finally, the RNN model is trained to the dataset using a single NVIDIA A100 GPU by minimising mean-squared error (MSE) loss over all sequence points and stress components using the adaptive moment estimation stochastic gradient descent (Adam) algorithm.

\section{Optimisation of network hyperparameters}
In order to understand the influence of the network parameters on the performance of the model, a convergence study was conducted varying parameter sizes from a baseline. The baseline parameters are listed in Table \ref{tab:baseline_params}, which lead to a model with approximately 1.1 million training parameters with \qty{70}{\percent} of these contained within the GRU block. As the model contains two mappings from initial microstructure, each of these will also be disabled to see if either is unnecessary. The dependence on the size of the training dataset will also be considered when training with smaller subsets. Each subset was produced by limiting the number of microstructures while keeping all load cases for a microstructure. In all cases the model was trained for 12 hours with a fixed learning rate of \num{1e-4}. Higher learning rates were initially used but this led to instabilities when training some of the models.

\begin{table}
    \caption{Baseline parameters for the RNN model described in Figure \ref{fig:network_diagram}.}
    \centering
    \begin{tabular}{rll}
        Parameter name & Baseline value & Value range \\\midrule
        Hidden state ($h_0$) & 256 & \numrange{64}{512} \\
        Microstructure input ($m_0$) & 32 & \numrange{8}{64} \\
        GRU layers ($n_l$) & 2 & \numrange{1}{16}
    \end{tabular}
    
    \label{tab:baseline_params}
\end{table}

\begin{figure}
  \centerline{\includegraphics{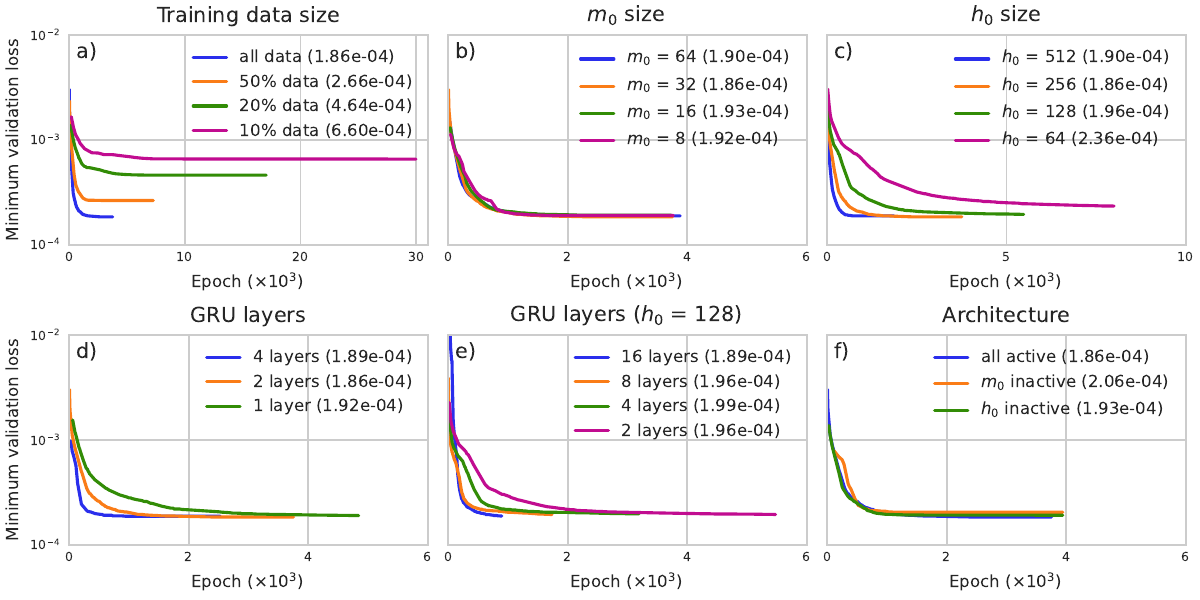}}
  \caption{Minimum validation loss curves for varying parameters from the baseline in Table \ref{tab:baseline_params}.}
  \label{fig:min_validation_loss}
\end{figure}

The training curves of the minimum validation error are shown in Figure \ref{fig:min_validation_loss} for each of the single-parameter sweeps and are also summarised in Table \ref{tab:losses}. There is a clear dependence of validation loss on the dataset size in Figure \ref{fig:min_validation_loss}a and we would expect this trend to continue if more training data was included. The final training loss values in Table \ref{tab:losses} show that the model does specialise to the training data, with lower training losses than for the validation set. This is also seen to cause overfitting beyond the minimum validation error, which is most prevalent in the models trained with less data.

\begin{table}
    \caption{Summary of model training losses for varying parameters sizes from the baseline values in Table \ref{tab:baseline_params}.}
    \centerline{
    \begin{tabular}{rllll|llll|llll}
                                    & \multicolumn{4}{l}{Dataset size}                                      & \multicolumn{4}{l}{$m_0$} & \multicolumn{4}{l}{$h_0$} \\
                                    & \qty{10}{\percent} & \qty{20}{\percent} & \qty{50}{\percent} & all    & 8 & 16 & 32 & 64          & 64 & 128 & 256 & 512 \\\midrule
        Minimum validation loss     & 6.60 & 4.64 & 2.66 & 1.86                                             & 1.92 & 1.93 & 1.86 & 1.90 & 2.36 & 1.96 & 1.86 & 1.90 \\
        Final validation loss       & 7.47 & 5.32 & 2.99 & 1.96                                             & 2.11 & 2.12 & 1.96 & 1.94 & 2.43 & 2.00 & 1.96 & 2.30 \\
        Final training loss         & 0.07 & 0.12 & 0.33 & 0.53                                             & 0.65 & 0.60 & 0.53 & 0.42 & 1.53 & 1.00 & 0.53 & 0.17 \\

    \rule{0pt}{4ex}                                   & \multicolumn{4}{l}{GRU layers}    & \multicolumn{4}{l}{GRU layers ($h_0 = 128$)}  & \multicolumn{4}{l}{Architecture} \\
                                    & 1 & 2 & 4 &                       & 2 & 4 & 8 & 16                                & all & $m_0$ off & $h_0$ off & \\\midrule
        Minimum validation loss     & 1.90 & 1.86 & 1.89 &              & 1.96 & 1.99 & 1.96 & 1.89                     & 1.86 & 2.06 & 1.93 & \\
        Final validation loss       & 1.95 & 1.96 & 2.06 &              & 2.00 & 2.02 & 2.02 & 1.93                     & 1.96 & 2.16 & 2.11 & \\
        Final training loss         & 0.84 & 0.53 & 0.34 &              & 1.00 & 0.78 & 0.85 & 0.80                     & 0.53 & 0.62 & 0.51 &
    \end{tabular}
    }
    \label{tab:losses}
\end{table}

Only small variations in the validation loss are observed for the models trained with the entire dataset (Figure \ref{fig:min_validation_loss}b-f). In most cases, the loss curve reaches a plateau in the range \numrange{1.85e-4}{2.00e-4} and only a small decrease in loss would be expected with further training. The complexity of the model influences the training behaviour, where models with fewer parameters (e.g. $h_0=64$) require more epochs to reach a plateau in loss. The complexity of the model is primarily controlled by the hidden state size and the number of layers in each recurrent block. We find that increasing the state size ($h_0$) above the baseline of \num{256} causes an increase in the minimum validation loss and worse overfitting past this minimum. However, this is not seen when increasing the number of GRU layers with a smaller state size ($h_0=128$).

\begin{figure}
  \centerline{\includegraphics{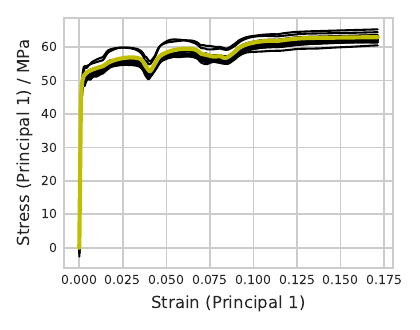}}
  \caption{Stress prediction for a single training point for each set of network parameters considered and the true output from the CP simulation (yellow).}
  \label{fig:single_stress_prediction}
\end{figure}

It is difficult to understand the behaviour of the model at all of the time steps using a single loss value. Figure \ref{fig:single_stress_prediction} shows the entire stress response for a single training data for the model trained with each set of parameters, and it can be seen that the small differences in validation loss lead to a spread of stress response after yield. Although, the predicted outputs do appear similar and shifted by a constant stress for the entire output range. Instabilities in the stress can also be seen during early plastic deformation, most likely due to the large change in gradient from elastic to plastic deformation. 

\begin{figure}
  \centerline{\includegraphics{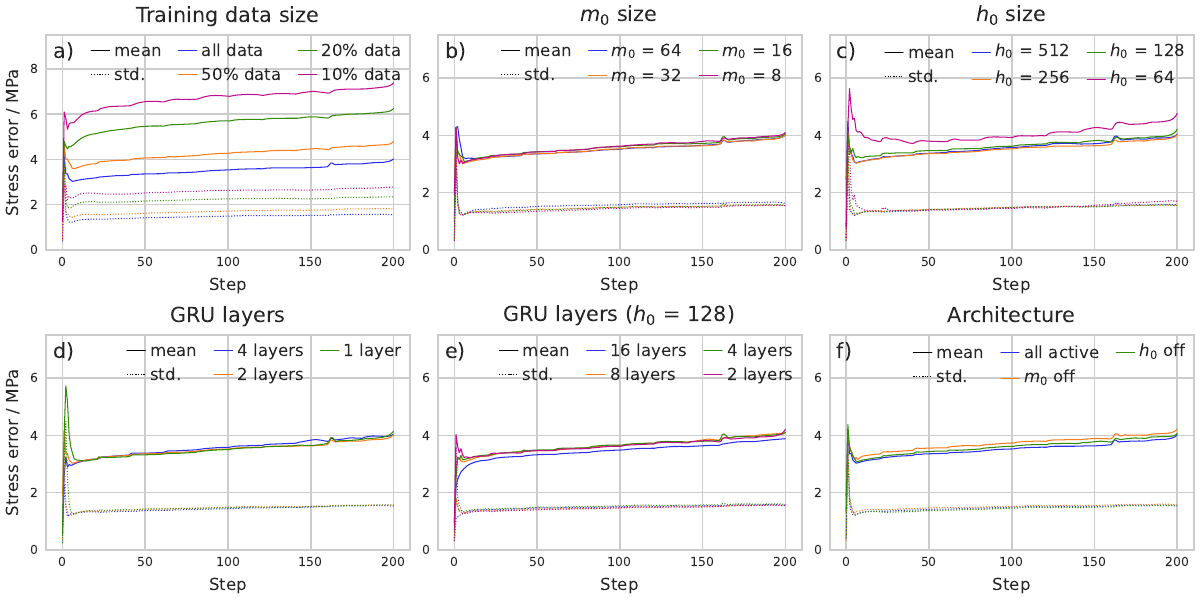}}
  \caption{Mean and standard deviation of stress error (Equation \ref{eqn:stressError}) over the validation set plotted over sequence step for each set of network parameters considered.}
  \label{fig:mean_stress_error_over_time}
\end{figure}

A stress error between the prediction of the surrogate model $\sigma_\text{model}$ and the true output $\sigma_\text{true}$ of the CP simulation is defined as,
\begin{equation}
\label{eqn:stressError}
\text{Stress error}=\lvert\lambda_\text{max}\{\sigma_\text{true}-\sigma_\text{model}\}\rvert,
\end{equation}
where $\lambda_\text{max}\{*\}$ denotes the matrix eigenvalue with largest absolute value. This error value represents the largest possible difference in stress which, due the random directions of the load paths, will not be aligned with a coordinate direction as with simple load paths. The stress error is plotted over time in Figure \ref{fig:mean_stress_error_over_time} for each of the parameter sweeps, with both the mean and standard deviation over the validation dataset shown. A similar progression of error over time is seen in all cases where the model is trained with the entire dataset, with the mean error seen to slowly increase with the time step but the standard deviation remains constant. The instability around yield is clearly seen as a peak in error for the early time steps and appears in almost all cases but is reduced when more layers are used in the GRU block and is removed for the case with 16 layers in Figure \ref{fig:mean_stress_error_over_time}e. Another smaller peak in error is seen at step 160, which corresponds to a strain rate change in the load case but it is unclear why a peak is not seen at the other strain rate changes.

\section{Validation against CP model}

\subsection{Independent variation of inputs}
The surrogate model has two groups of inputs that define the microstructure and the loading history independently and here we consider the influence of each input on the output stress. The training dataset has a heterogeneous structure in which each microstructure is simulated along many randomly selected load paths, with the number of load paths varying in the dataset. A similar structure is also present for load paths. The validation dataset does not consider the inputs separately, and all microstructures and load paths in the validation dataset do not appear in the training dataset. The validation dataset consists of pairs of microstructures and load paths which are both not present in the train data. Here we will consider cases when one of the microstructure or load case appear in the training data.

\begin{figure}
  \centerline{\includegraphics{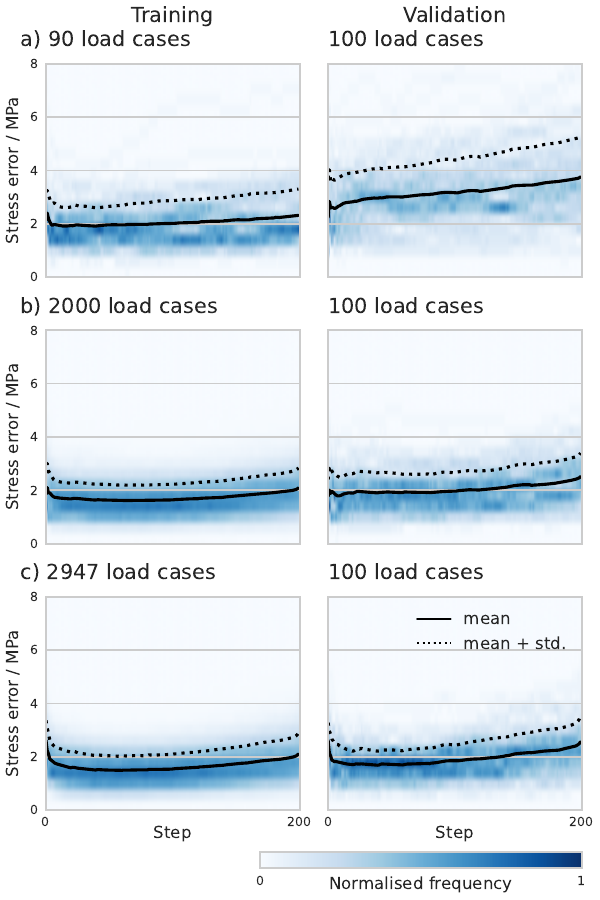}}
  \caption{Histograms over stress error (Equation \ref{eqn:stressError} plot) for each sequence (time) step for a fixed VE in each of a)-c) where a different number of load paths are included in the training set. The left column shows error in the simulations of the fixed VE from the training set and the right column shows the error of simulations for 100 unseen load paths for the same VE.}
  \label{fig:loadcase_verification}
\end{figure}

Considering first the case of fixed microstructures, Figure \ref{fig:loadcase_verification} shows the stress error (Equation \ref{eqn:stressError}) at each step of the output. In each plot, the distribution of the error is shown by the colourmap as well as mean and standard deviation. Part a) shows the case where \num{90} simulations with different load paths are included in the training dataset. The error of these training simulations is shown in the left plot and represents the minimum possible error that would be expected for unseen data. A new set of \num{100} simulations of the same microstructure but with unseen load paths are used for validation and the errors are shown in the right column of plots. The mean error in the validation simulations varies between \SIrange{2.5}{4}{\mega\pascal} and increases with the time step. The error is consistently higher than for the training data, suggesting that more training data could be beneficial.

In parts b) and c) of Figure \ref{fig:loadcase_verification} the stress error distributions are shown for microstructures with \num{2000} and \num{2947} load path simulations in the training dataset. In both cases, the validation is approximately \SI{2}{\mega\pascal} throughout and the spread in errors has decreased compared to the microstructure with \num{90} training points. There is a peak in error at the initial steps, which is due to the large change in gradient around yield. The mean error in the validation data is similar to that seen for the training data, suggesting that \num{2000} training load paths are sufficient to train the model for a single microstructure.

\begin{figure}
  \centerline{\includegraphics{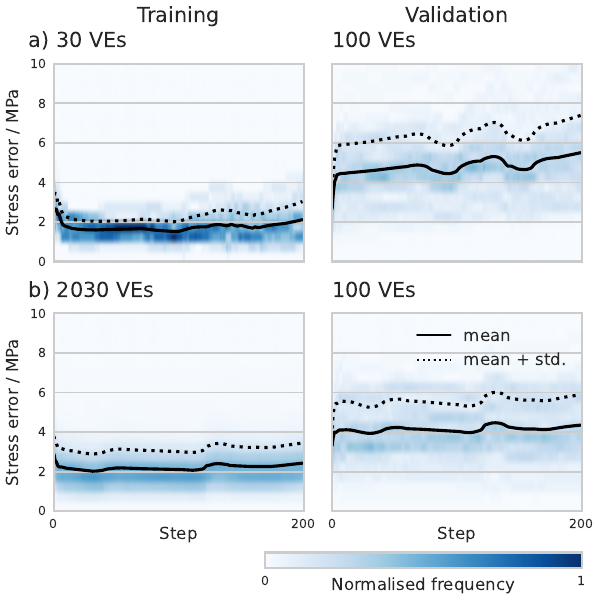}}
  \caption{Histograms over stress error (Equation \ref{eqn:stressError} plot) for each sequence (time) step for a fixed load path in each of a) and b) where a different number of VEs are included in the training set. The left column shows error in the simulations of the fixed load path from the training set and the right column shows the error of simulations for 100 unseen VEs for the same load path.}
  \label{fig:VE_verification}
\end{figure}

The opposite view will now be considered, keeping the load path fixed and varying the microstructure. Figure \ref{fig:VE_verification} shows the stress error distributions for this case, as before, training data is shown in the left column as a comparison to the unseen validation data in the right column. The errors in the training data are similar to those observed for fixed microstructure (Figure \ref{fig:loadcase_verification}), of approximately \SI{2}{\mega\pascal} with an initial peak at yield and a slight increase for later time steps. Part a) shows a load path where \num{30} simulations with different microstructures are included in the training dataset. The mean error varies in the range \SIrange{4}{6}{\mega\pascal}, with two troughs that are related to features in the fixed load path. Part b) shows a load path with \num{2030} training simulations. A decrease in mean error is seen compared to part a) but this is most prevalent at later time steps and the trend for larger error at later time steps is not seen. The error is approximately \SI{4}{\mega\pascal} throughout and has not converged to the error seen for the training data, suggesting that the training data is insufficient.

\subsection{Out-of-distribution inputs}
The training data generation process gives random coverage of input spaces of microstructure and loading condition but is not specialised to inputs that are of particular interest, such as idealised loading conditions and textured microstructures. These inputs are used to test the trained model against unseen data that will be important for specific applications where tailored training data may also be required. 

\begin{figure}
  \centerline{\includegraphics{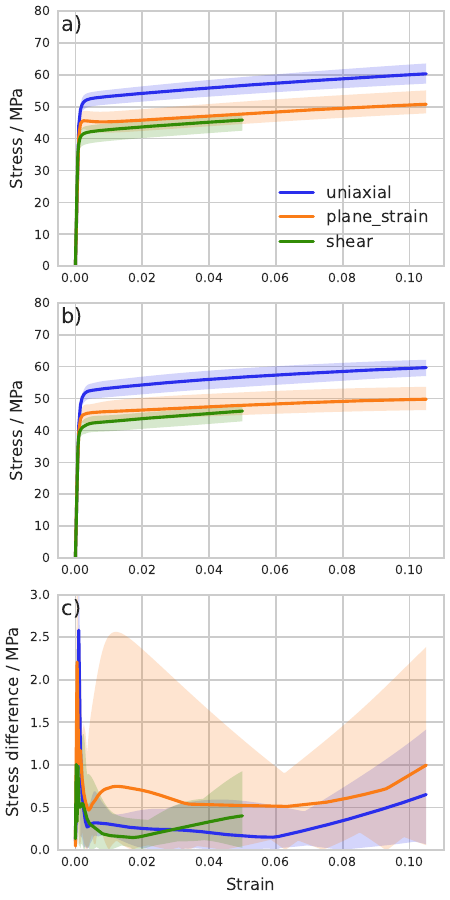}}
  \caption{Stress response of the a) CP model and b) surrogate model for idealised load paths that are not included in the training data. The solid line is the mean response of \num{10} simulations with different microstructures and the shaded area the min-max bounds. Part c) shows the difference in stress for calculated for each simulation.}
  \label{fig:out_of_distribution_loadcase}
\end{figure}

Uniaxial tension, plane strain and pure shear have been considered for the idealised load cases. Unlike the training set, these loading conditions include stress boundary conditions as well as conditions on the deformation gradient. The uniaxial case is defined as,
\begin{equation}
\label{eqn:uniaxial_loading}
\mathbf{F}=
\begin{pmatrix}
1.1 & 0 & 0 \\
0 & * & 0 \\
0 & 0 & * \\
\end{pmatrix},
\qquad\qquad
\mathbf{P}=
\begin{pmatrix}
* & * & * \\
* & 0 & * \\
* & * & 0 \\
\end{pmatrix},
\end{equation}
where $\mathbf{F}$ is the final deformation gradient, $\mathbf{P}$ is the first Piola–Kirchhoff stress for which the condition fulfilled throughout and * denotes components where the condition is applied on the other tensor. The plane strain load path is defined as,
\begin{equation}
\mathbf{F}=
\begin{pmatrix}
0.9 & 0 & 0 \\
0 & 1 & 0 \\
0 & 0 & * \\
\end{pmatrix},
\qquad\qquad
\mathbf{P}=
\begin{pmatrix}
* & * & * \\
* & * & * \\
* & * & 0 \\
\end{pmatrix},
\end{equation}
and the pure shear case as,
\begin{equation}
\mathbf{F}=
\begin{pmatrix}
1 & 0.1 & 0 \\
0.1 & 1 & 0 \\
0 & 0 & 1 \\
\end{pmatrix}.
\end{equation}
Results of \num{10} CP simulations with random microstructures are used to assess the surrogate model outputs of each of the load cases and are shown in Figure \ref{fig:out_of_distribution_loadcase}. Parts a) and b) shows the mean response of the \num{10} simulations for the CP and surrogate models respectively, with the shaded region indicating the maximum and minimum stress response. In each case the major loading direction is plotted: $\sigma_{11}$ for uniaxial, $\sigma_{33}$ for plane strain and $\sigma_{12}$ for shear. Good agreement is seen throughout each of the curves, with exception of the small peak in stress at yield for the plane strain case, which is not reproduced by the surrogate model. This is supported by the plot of stress difference in part c), which are generally below \SI{1}{\mega\pascal} and a maximum of \SI{2.5}{\mega\pascal} is seen around yield.

\begin{figure}
  \centerline{\includegraphics{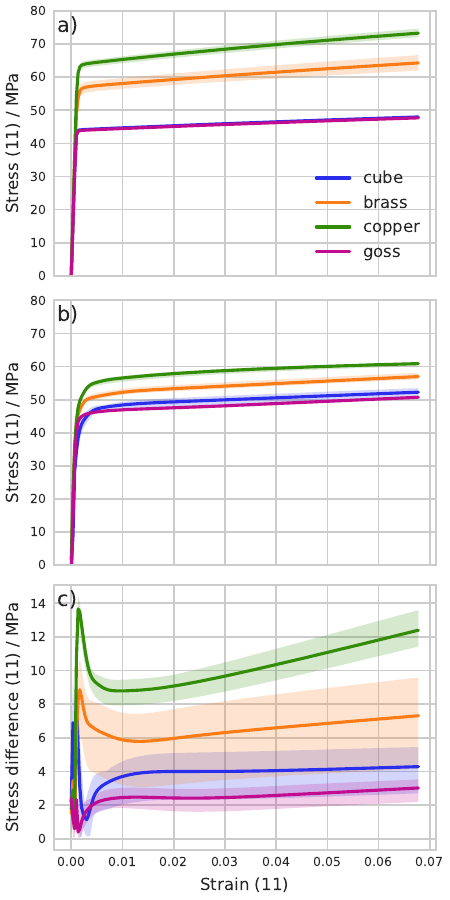}}
  \caption{Stress response of the a) CP model and b) surrogate model for textured microstructures loaded uiaxially that are not included in the training data. The solid line is the mean response of \num{10} simulations with a different sampling of orientations and the shaded area the min-max bounds. Part c) shows the difference in stress for calculated for each simulation.}
  \label{fig:out_of_distribution_texture}
\end{figure}

Textured microstructures are considered by sampling orientations from an orientation distribution function (ODF) defined to represent common texture components. MTEX \citep{bachmann_texture_2010}, an open source library for texture analysis, is used to create and sample ODFs. Textured ODFs are defined by a Gaussian-like kernel with a halfwidth of \ang{7.5} placed at the orientation that defines the texture component. Table \ref{tab:texture_comps} shows the texture components considered. The \num{20} sampled orientations are then combined with a set of random Voronoi seeds to define a VE. \num{10} VEs were generated for each texture and CP simulations produced by loading in uniaxial tension. Results of these simulations and outputs from the surrogate model are shown in Figure \ref{fig:out_of_distribution_texture}. The surrogate response is not as closely matched to the CP simulation as with the idealised load cases and the stress responses all appear to be shifted to a mean response of randomly chosen orientations. However, the progression of strength of the texture components is correct, with brass seen to have the highest and Goss the lowest stress response in both CP and surrogate. 

\begin{table}
    \caption{Texture components considered for microstructure edge cases. \citep{Kocks2000}}
    \centering
    \begin{tabular}{r|l}
        Texture & Orientation ($\phi_1$, $\Phi$, $\phi_2$)\\
        Goss & \ang{0}, \ang{45}, \ang{0}\\
        Brass & \ang{35}, \ang{45}, \ang{0}\\
        Cube & \ang{0}, \ang{0}, \ang{0}\\
        Copper & \ang{90}, \ang{35}, \ang{45}\\
    \end{tabular}
    \label{tab:texture_comps}
\end{table}

\section{Embedding in larger scale simulations}
The aim of developing this surrogate model is to allow the mechanical response of a CP model to be used in applications that would otherwise be prohibitively computationally expensive, such as multilevel finite element method (FE\textsuperscript{2}) \citep{feyel_multilevel_2003} and sampling based uncertainty propagation \citep{Tran2021}. Here we will demonstrate the surrogate in an FE\textsuperscript{2} application by embedding it as the constitutive model of a larger scale simulation. The large-scale model is constructed using an FFT algorithm similar to the CPFFT model used to generate the training data \citep{Eisenlohr2013,Shanthraj2015,de_geus_finite_2017}. This algorithm allows solving for quasi-static stress equilibrium of a non-linear material without a stress Jacobian or Newton-Raphson incremental update. A hybrid CPU-GPU model is implemented with the GPU used to calculate the RNN constitutive response and a residual deformation field via FFT convolution, which is closer to satisfying equilibrium. The residual is passed to the CPU where the N-GMRES implemented in PETSC \citep{balay_petsctao_2025} is used to update the local deformation field and this loop continues until the relative error in satisfying stress equilibrium is below \num{5e-3}. This approach uses the GPU for efficient NN and FFT calculations but minimises slow copying operations between GPU and CPU memory, and the large hidden state vector, stored at each grid point, remains in GPU memory throughout.

\begin{figure}
  \centerline{\includegraphics{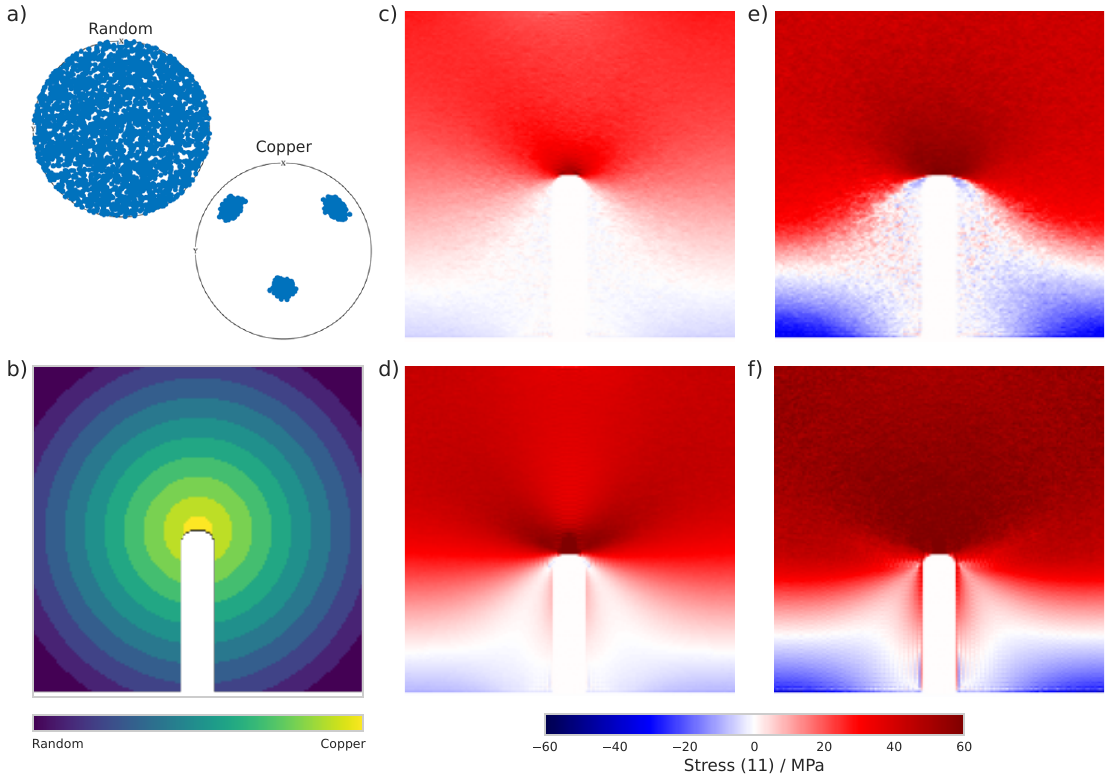}}
  \caption{Results of the RNN model embedded in a larger scale simulation with varying microstructure and large blunted crack feature. Part a) shows the two extremes of textures that are mixed to produce a steady variation of microstructure shown in part b). Parts a) and e) show the stress response of the embedded RNN model at low (\num{0.001}) and high strain (\num{0.01}) respectively, and parts d) and e) show the same but for a Taylor (isostrain) homogenisation model.}
  \label{fig:fft_solver_example}
\end{figure}

The model is used to calculate the stress distribution around a large blunted crack embedded in a region of material with variable crystallographic texture. Figure \ref{fig:fft_solver_example}b shows a slice through the \numproduct{151x151x15} domain, where each banded region has orientations sampled from a progression of ODFs from strong copper texture to random texture. Each point in the simulation is represented by an individual \num{20} grain Voronoi tessellation, with seed positions randomly sampled and each point having a unique sampling of orientations. The slot shown in the figure is an elastic region with a vanishingly low modulus ($\mathbf{C}_{11}$=\SI{0.1}{\mega\pascal} or 6 orders of magnitude lower than the CP model, Table \ref{tab:sim_params_damask}) which reproduced the behaviour of a void in the model \citep{Maiti2018}. As these models simulate periodic deformation, a two voxel thick layer of the elastic material is also added to the face of the volume with normal in the y-direction (lower edge in the figure), creating a free surface on this face and decoupling the two faces. Uniaxial loading is applied to the volume in the x-direction (horizontal in the figure) in \num{10} equal steps to a final average macroscopic strain of \num{0.01}. Results from the first and final increments are shown in parts c and e of Figure \ref{fig:fft_solver_example}, where a stress concentration is seen at the top of the slot which decays with distance and the local variation caused by small differences in microstructure sampling.

A multiscale DAMASK simulation is used as a comparison to the surrogate model. At each grid point of the simulation, the mechanical response is determined from the homogenised response of multiple orientations, using an iso-strain (Taylor) assumption over each ensemble. The volume with graded texture is reproduced in DAMASK, with the same 20 orientations defined at each point. The same macroscopic loading is applied and the results from the first and final increments are included in Figure \ref{fig:fft_solver_example}d and f. A similar spatial distribution in stress is seen between the two models, indicating the stress equilibrium calculation in the FE\textsuperscript{2} surrogate model works as intended. Sharper gradients in stress around the stress concentration are seen in the DAMASK model, which is typical of sharper stress transitions seen in Taylor models compared to full-field simulations. Despite the simpler isostrain assumption in the Taylor model, the surrogate model example had a significantly lower runtime. The DAMASK simulation used 56-cores and ran in \num{62}~minutes, for which the majority of this time was used for constitutive model solves. A single CPU core and a GPU were used to run the surrogate model and it ran in \num{3}~minutes but only approximately 4~seconds of this was constitutive model solves.

\section{Discussion}
\begin{figure}
  \centerline{\includegraphics{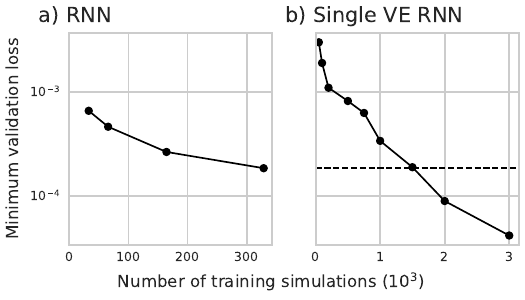}}
  \caption{Plot of minimum validation loss for varying training dataset sizes for the a) full microstructure dependent RNN model and b) an RNN model trained with a single fixed microstructure.}
  \label{fig:min_validation_loss_vs_data}
\end{figure}

We have demonstrated the use of an RNN as a microstructure sensitive surrogate model for the homogenised mechanical response of a full-field CP model. The model is defined over two complex inputs: a representation of 3D microstructure by Voronoi tessellation and a sequence of deformation steps. This parameter space required a large dataset of CP simulations, the generation of which required significant compute resources. However, this compute is only required once and can be easily parallelised over many thousands of cores of a compute cluster, unlike for the eventual use cases of FE\textsuperscript{2} simulation or sequential sampling based uncertainty quantification. The model was shown to have greater accuracy in capturing load path dependence when compared with microstructure dependence, suggesting that more VEs should be added to the training dataset to increase accuracy. A continual decrease in validation loss was observed with the size of the dataset (Figure \ref{fig:min_validation_loss}a) and this trend is summarised in Figure \ref{fig:min_validation_loss_vs_data}a. Figure \ref{fig:min_validation_loss_vs_data}b then shows the loss of a simplified RNN model with fixed microstructure for various dataset sizes (using the data from Figure \ref{fig:loadcase_verification}). Beyond \num{1500} training points, the loss is lower for the simplified model and this can be used as a limit for the number of load paths for each VE added to the dataset to efficiently improve accuracy. Although the input data used here was generated randomly, active learning approaches have been used for similar models, guided by larger scale simulations for load path placement \citep{zhang_iterated_2024} in an FE\textsuperscript{2} use case or more generally by finding areas of the input parameter space with highest error \citep{zanisi_efficient_2024}. The microstructure data generation could also be guided by the properties of a known material, such as grain shape and texture, to reduce the required data, whereas here we attempted to capture the variance across every possible VE in our training dataset.

RNNs have been shown to model time sequential data from history-dependent deformation well, with the internal variable representation of the physical models interpreted well by non-linear updates to a hidden state \citep{abueidda_deep_2021, bonatti_cp-fft_2022, liu_learning_2023}. Extending the RNN model to include variation of the hidden state to represent the initial microstructure has required a large amount of training data and large hidden state sizes. Due to the volume of training, we used a GRU architecture for the RNN portion of the model as efficient implementations are available, although each epoch took \SI{50}{\second} for the 16-layer model and several thousand were required to train the model to the minimum loss. RNN architectures tailored towards deformation models have been developed, such as recurrent neural operators \citep{liu_learning_2023} and linearised minimal state cell \citep{bonatti_importance_2022, bonatti_cp-fft_2022} but are trained with an order of magnitude less data. These are implemented as cell update with a loop over the sequence length and when this change was made to the GRU model used here each epoch took 5.5 times longer, despite being an otherwise identical model. RNNs must typically be calculated sequentially due to the non-linear updates, meaning generalisation to long sequence data is difficult. Linear RNNs have been proposed \citep{martin_parallelizing_2018, feng_were_2024}, which are more efficient as the sequence updates can be calculated in parallel. However, models of this type did not represent our data well, with a minimum validation loss of \num{7e-4} for the fixed state model shown in Figure \ref{fig:min_validation_loss_vs_data}b compared to \num{4e-5} for the non-linear model. Due to the popularity of large language models, there are many options for modelling sequence data but attention has been on complex transformer architectures. Linear state space models are still being investigated, such as linear RNNs or novel modifications seen in the mamba model \citep{gu_mamba_2024} with mechanisms to select important state data to reproduce the non-linear behaviour of the reset gate in a GRU.

\begin{figure}
  \centerline{\includegraphics{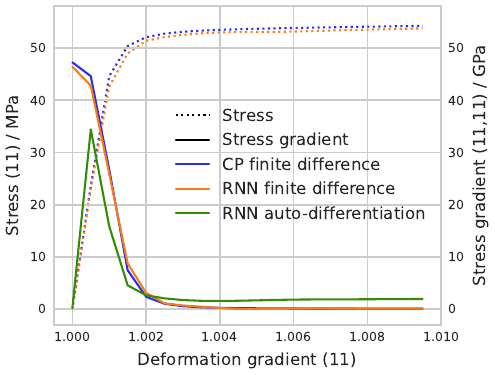}}
  \caption{Gradient response of the CP and RNN model for uniaxial loading calculated by finite differences and auto differentiation of the RNN model.}
  \label{fig:gradient_comparison}
\end{figure}

The value of the surrogate model has been demonstrated by integrating it into a larger scale simulation as the constitutive model, using an FFT algorithm to solve for stress equilibrium. Typically, the finite element method (FEM) is used for engineering simulation but we found issues with integrating the surrogate with the non-linear iterative Newton-Raphson (NR) algorithm when solving static equilibrium. These methods require a stress and stress Jacobian from the constitutive model, which can be calculated using auto-differentiation. However, we found this gradient to be inconsistent for the initial output, shown in Figure \ref{fig:gradient_comparison} for a uniaxial loading condition (Equation \ref{eqn:uniaxial_loading}). The gradient was also calculated directly from the stress using finite differences using both the CP and RNN output, which give similar results. It would appear that the instantaneous gradient is not well defined at $t=0$, which will be approached as the NR algorithm finds a consistent solution. RNN deformation models have been used in FEM previously but have used explicit updates \citep{liu_learning_2023, bonatti_cp-fft_2022, zhang_iterated_2024, tang_crystal_2025} without non-linear iterations to find static equilibrium. The FFT algorithm does not require a gradient and each non-linear iteration is calculated for the entire time step, not a decreasing fraction as with NR, and shows RNN deformation models can be used in static equilibrium calculations.

\section{Conclusion}
In this work, we developed a microstructure-sensitive RNN surrogate for full-field crystal plasticity, capable of linking 3D microstructure and deformation history to homogenised stress response with high efficiency. The model aimed to generalise to any loading path and microstructure by randomly sampling both input parameters, and was demonstrated to predict unseen validation data with good accuracy throughout the sequence of stress output. We demonstrate the model by embedding it as the constitutive model of a larger simulation, where it could be used to solve for static equilibrium with varied microstructure input. While performance was best for load path generalisation, improvements are still needed for out-of-distribution microstructures, motivating future efforts in refining the model architecture and active or targetted training data generation.

\section*{Acknowledgements}
This work was part funded by the FARSCAPE project, a collaboration between
UKAEA and the UKRI-STFC Hartree Centre, and by the EPSRC Energy programme (grant number EP/W006839/1).

\newpage
\bibliographystyle{elsarticle-harv} 
\bibliography{main.bib}

\end{document}